\documentclass{ws-procs961x669}            
\begin{document}
\title{Cosmological singularities in Eddington-inspired-Born-Infeld theory and its possible extension}

\author{Che-Yu Chen$^{1,2}$, Mariam Bouhmadi-L\'{o}pez$^{3,4,5,6}$ and Pisin Chen$^{1,2,7}$}

\address{${}^1$Department of Physics, National Taiwan University, Taipei, Taiwan 10617\\
${}^2$LeCosPA, National Taiwan University, Taipei, Taiwan 10617\\
${}^3$Departamento de F\'{i}sica, Universidade da Beira Interior, 6200 Covilh\~a, Portugal\\
${}^4$Centro de Matem\'atica e Aplica\c{c}\~oes da Universidade da Beira Interior (CMA-UBI) 6200 Covilh\~a, Portugal\\
${}^5$Department of Theoretical Physics, University of the Basque Country UPV/EHU, P.O. Box 644, 48080 Bilbao, Spain\\
${}^6$IKERBASQUE, Basque Foundation for Science, 48011, Bilbao, Spain\\
${}^7$Kavli Institute for Particle Astrophysics and Cosmology, SLAC National Accelerator Laboratory, Stanford University, Stanford, CA 94305, U.S.A.\\
E-mail: b97202056@gmail.com, mbl@ubi.pt, pisinchen@phys.ntu.edu.tw\\
}

\begin{abstract}
The Eddington-inspired-Born-Infeld (EiBI) gravity, which is formulated within the Palatini formalism, is characterized by its ability to cure the big bang singularity in the very beginning of the Universe. We further analyze the EiBI phantom model, and investigate the possible avoidance or alleviation of other dark energy related singularities. We find that except for the big rip singularity and little rip event, most of the cosmological singularities of interest can be partially alleviated in this model. Furthermore, we generalize the EiBI theory by adding a pure trace term to the determinant of the action. This amendment is the most general rank-two tensor composed of up to first order of the Riemann curvature. We find that this model allows the occurrence of primitive bounces and some smoother singularities than that of big bang. Most interestingly, for certain parameter space, the big bang singularity can be followed naturally by an inflationary stage in a radiation dominated universe.
\end{abstract}

\keywords{modified theories of gravity, early time universe, Eddington-inspired-Born-Infeld theories, cosmic singularities}

\bodymatter

\section{Introduction}\label{aba:sec1}
In 1934 Born and Infeld proposed a non-linear action for classical electrodynamics, which is characterized by its success in solving the divergence of the self-energy of point-like charges \cite{Born:1934gh}. Since their proposal, modified theories of gravity with a Born-Infeld-inspired action have received much attention (initiated in Ref.~\citenum{Deser:1998rj}). These theories not only maintain the properties of general relativity (GR) for small curvatures, but provide various interesting deviations from GR at high curvature regimes. Some of these amendments are hoped to shed some light on smoothing the singularities in GR. For example, recently a theory dubbed Eddington-inspired-Born-Infeld theory (EiBI) \cite{Banados:2010ix} has been studied from both astrophysical and cosmological points of view. The EiBI theory is shown to be able to cure the big bang singularity for a radiation dominated universe. Therefore, in Refs.~\citenum{Bouhmadi-Lopez:2013lha} and \citenum{Bouhmadi-Lopez:2014jfa} we carry out a thorough analysis on the possible avoidance of other cosmological singularities such as big rip, sudden, big freeze, type IV singularities, and little rip event in the EiBI theory. We find that though the big rip singularity and little rip can not be avoided, most of the cosmological singularities can be alleviated in the EiBI theory \cite{Bouhmadi-Lopez:2014jfa}.

In Ref.~\citenum{Chen:2015eha}, we extend the EiBI theory by adding a trace term into the determinantal action. It corresponds to the most general action constructed from a rank two tensor that contains up to first order terms in curvature. This model can provide smooth bouncing solutions which were not allowed in the EiBI model for the same EiBI coupling. Most interestingly, for a radiation filled universe there are some regions of the parameter space that can naturally lead to a de Sitter inflationary stage without any exotic matter field.

\section{The EiBI cosmology}
We start reviewing the EiBI model whose action in terms of the metric $g_{\mu\nu}$ and the connection $\Gamma^{\alpha}_{\mu\nu}$
reads \cite{Banados:2010ix} 
\begin{equation}
\mathcal{S}_{\textrm{EiBI}}(g,\Gamma,\Psi)=\frac{2}{\kappa}\int d^4x\left[\sqrt{|g_{\mu\nu}+\kappa R_{\mu\nu}(\Gamma)|}-\lambda\sqrt{|g|}\right]+\mathcal{S}_\textrm{m} (g,\Psi).
\label{action}
\end{equation} 
The theory is formulated within the Palatini approach and the connection is assumed to be torsionless.\footnote{See our work Ref.~\citenum{Bouhmadi-Lopez:2014tna} where torsion is taken into account} Within this setup, the connection $\Gamma^{\alpha}_{\mu\nu}$ and the metric $g_{\mu\nu}$ are treated as independent variables. The parameter $\kappa$ is a constant with inverse dimensions to that of a cosmological constant ($8\pi{G}=1$ and $c=1$), $\lambda$ is a dimensionless constant parametrizing the cosmological constant and $\mathcal{S}_\textrm{m} (g,\Psi)$ stands for the matter Lagrangian in which matter is assumed to be coupled covariantly to the metric $g$ only. 

For a FLRW universe filled with a perfect fluid with energy density $\rho$ and pressure $p$, the Friedmann equation reads \cite{Avelino:2012ue}
\begin{equation}
\bar{H}^2=\frac{8}{3}\frac{[\bar\rho+3\bar p-2+2\sqrt{(1+\bar\rho)(1-\bar p)^3}](1+\bar\rho)(1-\bar p)^2}{[(1-\bar p)(4+\bar\rho-3\bar p)+3\frac{d\bar p}{d\bar\rho}(1+\bar\rho)(\bar\rho+\bar p)]^2},
\label{field equation}
\end{equation}
where $\bar H\equiv\sqrt{\kappa}H$, $H$ is the Hubble parameter as defined from the physical metric, $\bar\rho=\kappa\rho$, $\bar p=\kappa p$. For simplicity, we will also use a dimensionless cosmic time: $\bar t\equiv t/\sqrt{\kappa}$ where $t$ corresponds to the cosmic time of the physical metric $g_{\mu\nu}$. It can be easily verified that the big bang singularity can be avoided in this theory for a radiation dominated universe \cite{Banados:2010ix}. In general a universe filled with a perfect fluid with a constant and positive equation of state $w$ bounces in the past for $\kappa<0$ or has a loitering behavior in the infinite past for $\kappa>0$ \cite{Scargill:2012kg}.

Aside, the auxiliary metric $q_{\mu\nu}$ which is compatible with the connection is\cite{Banados:2010ix}:
\begin{equation}
q_{\mu\nu}dx^\mu dx^\nu=-U(t)dt^2+a^2(t)V(t)(dx^2+dy^2+dz^2),
\end{equation}
where 
\begin{equation}
U=\sqrt{\frac{(1-\bar p)^3}{1+\bar\rho}},\ \ \ \ \ V=\sqrt{(1+\bar\rho)(1-\bar p)}.\label{V}
\end{equation}
From the auxiliary metric $q_{\mu\nu}$ we can define as well a rescaled dimensionless auxiliary Hubble parameter $\bar{H}_q$ as follows
\begin{eqnarray}
\bar{H}_q=\sqrt{\kappa}\frac{1}{\tilde{a}}\frac{d\tilde{a}}{d\tilde t}=\frac{1}{\sqrt{U}}\frac{d}{d\bar{t}}\ln(a\sqrt{V}),
\label{defHq}
\end{eqnarray}
where $\tilde a\equiv\sqrt{V}a$ and $d\tilde t\equiv\sqrt{U}dt$. Then we have
\begin{equation}
{\bar H_q}^2=\frac{1}{3}+\frac{\bar\rho+3\bar p-2}{6\sqrt{(1+\bar\rho)(1-\bar p)^3}}.
\label{HHq}
\end{equation}

\subsection{The EiBI phantom model}
We then analyze the possible avoidance of phantom dark energy related singularities in the EiBI setup. These singularities are characterized by possible divergence of the Hubble parameter and its cosmic time derivatives at some finite cosmic time (See Ref.~\citenum{Bouhmadi-Lopez:2014jfa} and references therein). Note that we have two possible Hubble rate: $H$ defined by $g_{\mu\nu}$ and $H_q$ defined by $q_{\mu\nu}$.

For the sake of completeness, we assume the Universe is filled with radiation, dark and baryonic matter, and dark energy:
\begin{equation}
\bar\rho=\bar\rho_r+\bar\rho_m+\bar\rho_{de},\ \ \ \ \ \bar p=\frac{1}{3}\bar\rho_r+\bar p_{de}(\bar\rho_{de}).\
\label{content}
\end{equation}
Note that $p_{de}(\bar\rho_{de})$ means that the equation of state of dark energy is purely a function of the dark energy density.

To analyze the big rip singularities, we assume a phantom dark energy with constant equation state $w<-1$ to be the component of dark energy. For the big freeze, sudden, and type IV singularities, we regard the phantom Generalized Chaplygin Gas (pGCG) as the dark energy component in this model \cite{BouhmadiLopez:2004me,BouhmadiLopez:2007qb}. Its equation of state takes the form:
\begin{equation}
\bar p_{de}=-\frac{A}{(\bar\rho_{de})^{\alpha}},
\label{pgcg EOM}
\end{equation}
where $\alpha$ and $A>0$ are two dimensionless constants. The energy density in terms of the scale factor for $\alpha>-1$ is \cite{BouhmadiLopez:2004me,BouhmadiLopez:2007qb}:
\begin{equation}
\bar\rho_{de}=\bar\rho_{de0}\left[\frac{1-\left(\frac{a_\textrm{min}}{a}\right)^{3(1+\alpha)}}{1-{a_\textrm{min}}^{3(1+\alpha)}}\right]^{\frac{1}{1+\alpha}},
\label{rpp}
\end{equation}
where $a_\textrm{min}$ is the scale factor corresponding to the past singularity. A subscript $0$ stands for quantities evaluated today. The phantom energy of this kind will drive past sudden or type IV singularities in GR \cite{BouhmadiLopez:2004me,BouhmadiLopez:2007qb}. On the other hand, if $\alpha<-1$, the energy density of the pGCG, which drives a future big freeze singularity in GR, reads \cite{BouhmadiLopez:2004me,BouhmadiLopez:2007qb}:
\begin{equation}
\bar\rho_{de}=\bar\rho_{de0}\left[\frac{1-\left(\frac{a_\textrm{max}}{a}\right)^{3(1+\alpha)}}{1-{a_\textrm{max}}^{3(1+\alpha)}}\right]^{\frac{1}{1+\alpha}},
\label{rfs}
\end{equation}
where $a_\textrm{max}$ is the scale factor corresponding to the future singularity. For the little rip event, we consider the simplest and the most common-used dark energy equation of state driving the little rip in GR (See Ref.~\citenum{Albarran:2016ewi} and references therein)
\begin{equation}
\bar p_{de}=-\bar\rho_{de}-B\sqrt{\bar\rho_{de}}, 
\label{eqstatelr}
\end{equation}
where $B$ is a positive dimensionless constant. In each cases, we not only consider the asymptotic behaviors of the physical metric, but also consider the asymptotic behaviors of the auxiliary metric by analyzing the behaviors of $H_q$ and its derivatives with respect to $\tilde{t}$ near the singular events. The results are summarized in Table.~\ref{aba:tbl1} \cite{Bouhmadi-Lopez:2014jfa}.

\begin{table}
\tbl{This table summarizes how the asymptotic behaviour of a universe near the singularities in GR is altered in the EiBI theory when the Universe is filled with matter, radiation as well as phantom energy. The row labelled by (1) corresponds to $-1/3<\alpha<0$ or $-1<\alpha<-2/3$, and where $\alpha$ cannot be written as $-1/(n+2)$ or $-n/(n+1)$, with $n$ being a natural number. If $\alpha=-1/(n+2)$ ($-1/3\le\alpha<0$ naturally) which is labelled by (2), there is no singularity while the Universe starts to expand from a finite size at a finite cosmic time. Note that it is possible for the Universe to start from a loitering phase of the physical metric instead of a past singularities, as long as the total pressure reaches the value $\bar p=1$ at some particular scale factor $a_b$ such that $a_b>a_{\textrm{min}}$, and it corresponds to a past big bang singularity of the auxiliary metric.}
{\begin{tabular}{@{}|c|c|c|@{}}
\hline
   Singularity in GR & EiBI physical metric & EiBI auxiliary metric\\
  \hline\hline 
   Big Rip & Big Rip & expanding de-Sitter \\ 
  \hline
   past Sudden & past Type IV ($0<\alpha\leq2$) & contracting de-Sitter \\
   \cline{2-2}
   ($\alpha>0$)&past Sudden ($\alpha>2$)& \\
  \hline
   future Big Freeze &future Big Freeze ($-3<\alpha<-1$) & expanding de-Sitter \\
   \cline{2-2}
   ($\alpha<-1$)&future Type IV ($\alpha=-3$)&\\
   \cline{2-2}
   &future Sudden ($\alpha<-3$)&\\
  \hline
   past Type IV&past Sudden ($-2/3<\alpha<-1/3$)&past Type IV\\
   \cline{2-2}
   ($-1<\alpha<0$)&(1)past Type IV&\\
   \cline{2-3}
   ($\alpha\neq-n/(n+1)$)&(2)finite past without singularity&finite past without singularity\\
   \cline{2-3}
   & past loitering effect ($a_b>a_\textrm{min}$)& Big Bang\\
  \hline
  finite past without singularity&finite past without singularity&finite past without singularity\\
  \cline{2-3}
  ($\alpha=-n/(n+1)$)& past loitering effect ($a_b>a_\textrm{min}$)& Big Bang\\
  ($-1<\alpha<0$)&&\\
  \hline
  Little Rip & Little Rip & expanding de-Sitter\\
\hline
\end{tabular}
}
\label{aba:tbl1}
\end{table}

Besides, we also use a cosmographic approach to constrain the parameters present in our model \cite{Bouhmadi-Lopez:2014jfa}. As a result, it turns out that the cosmographic analyses pick up the physical region preferring  the occurrence of a type IV singularity in the finite past or the loitering effect in an infinite past. While it is necessary to impose more conditions to forecast the future doomsdays of the Universe in this model \cite{Bouhmadi-Lopez:2014jfa}.

\section{Modified EiBI theory - A pure trace term}
In Ref.~\citenum{Chen:2015eha} we add a pure trace term, which takes the form of $g_{\mu\nu}R$, to the EiBI determinantal Lagrangian. The action becomes
\begin{equation}
\mathcal{S}=\frac{2}{\kappa}\int d^4x\Big[\sqrt{|g_{\mu\nu}+\kappa[\alpha R_{\mu\nu}(\Gamma)+\beta g_{\mu\nu}R]|}-\lambda\sqrt{-g}\Big]+S_m.
\label{action}
\end{equation}
The dimensionless constants $\alpha$ and $\beta$ should satisfy $\alpha+4\beta=1$ to ensure the recovering of Einstein GR at the low curvature limit \cite{Chen:2015eha}.

The field equations with respect to $g_{\mu\nu}$ and the connection are 
\begin{eqnarray}
&&\frac{\sqrt{-q}}{\sqrt{-g}}q^{\mu\nu}(1+\kappa\beta R)-\lambda g^{\mu\nu}-\frac{\sqrt{-q}}{\sqrt{-g}}\kappa\beta q^{\alpha\beta}g_{\alpha\beta}g^{\mu\rho}g^{\nu\sigma}R_{\rho\sigma}=-\kappa T^{\mu\nu},\\
\label{eq1}
&&\nabla_\nu\Big[\sqrt{-q}(\alpha q^{\mu\nu}+\beta q^{\alpha\beta}g_{\alpha\beta}g^{\mu\nu})\Big]=0,
\label{eq2}
\end{eqnarray}
respectively, where $q_{\mu\nu}\equiv g_{\mu\nu}+\kappa\alpha R_{\mu\nu}(\Gamma)+\kappa\beta g_{\mu\nu}R$ and $q^{\mu\nu}$ is its inverse. Note that the covariant derivative $\nabla_\nu$ is defined through the affine connection.

In Ref.~\citenum{Chen:2015eha} we consider a flat FLRW universe filled with radiation in this type of model. To analyze the behavior of the cosmological solutions, we rewrite the field equations in an algebraic form and express the energy density and Hubble rate using a single variable $x$. A parametrized Hubble function can be obtained \cite{Chen:2015eha}. The cosmological solutions in the presence of the trace term, i.e, $\beta\neq0$, are summarized in Table.~\ref{aba:tbl2}. We highlight the following interesting results concerning this model: \cite{Chen:2015eha}
\begin{romanlist}[(iii)]
\item Bouncing solutions which were not allowed in the EiBI model for the same EiBI coupling $(\kappa>0)$
\item A de Sitter inflationary stage after the big bang singularity when $\beta\lesssim 1$
\item A quasi-sudden singularity where $dH/dt$ becomes very large but finite at a finite cosmic time
\end{romanlist}

\begin{table}
\tbl{This table summarizes how the big bang singularity in GR is altered in the modified EiBI theory for a radiation dominated universe. If $\kappa<0$, the big bang is substituted by a bounce except for the regions of the parameter space $0<\beta\leq1/4$ where the big bang is still present. If $\kappa>0$, the big bang singularity can be altered by a loitering effect, a bounce, what we named a quasi-sudden singularity, or a big freeze singularity in the past. However, for $1/4\leq\beta<1$, the big bang singularity exists. Furthermore, the big bang singularity may be followed by a de Sitter inflationary stage for $\beta\lesssim 1$.}
{\begin{tabular}{@{}|c|c|c|@{}}
\hline 
    & $\kappa>0$ & $\kappa<0$ \\
  \hline\hline
   $\beta<0$ &  &  \\ 
  \cline{1-1}
   $\beta=0$ & past loitering effect& bounce \\
   (EiBI theory)& & \\
  \hline\ 
   $0<\beta<\beta_{\star}$ & bounce &  \\
   \cline{1-2}
   $\beta=\beta_{\star}$&past quasi-sudden singularity & \\
   \cline{1-2}
   $\beta_{\star}<\beta<1/4$& past big freeze singularity& big bang singularity\\
  \cline{1-2}
   $\beta=1/4$ && \\
   Palatini $R^2$ theory&big bang singularity&\\
   \cline{1-1}\cline{3-3}
   $1/4<\beta<1$&&\\
   \cline{1-2}
   $\beta\lesssim 1$&big bang singularity$+$de Sitter&bounce\\
   \cline{1-2}
   $\beta\geq 1$& past loitering effect&\\
  \hline
\end{tabular}
}
\label{aba:tbl2}
\end{table}

\section{Summary}
We give a summarized analysis on the avoidance of dark energy related singularities in EiBI type of theories by investigating the asymptotic behaviors of the Hubble rate and its cosmic time derivatives defined by the physical metric $g_{\mu\nu}$, which is coupled to matter, and by the auxiliary metric $q_{\mu\nu}$ compatible with the physical connection. For the physical metric $g_{\mu\nu}$ we find that though the big rip singularity and the little rip event driven by phantom dark energy are not cured in the EiBI theory, this theory to some extent smoothes the other phantom dark energy related singularities present in GR \cite{Bouhmadi-Lopez:2014jfa}. The behavior of the auxiliary metric $q_{\mu\nu}$ is even more regular (See Table.~\ref{aba:tbl1} for a summary). The fate of a bound structure near the singularities in the EiBI theory and some constraints on the model from a cosmographic approach are also analyzed in Ref.~\citenum{Bouhmadi-Lopez:2014jfa}.

Furthermore, a modification of the EiBI theory by adding a pure trace term into the determinantal Lagrangian and its cosmological solutions are investigated \cite{Chen:2015eha}. We assume a homogeneous and isotropic universe filled with radiation, and analyze the behaviors of the cosmological solutions using a parametric Friedmann equation. We summarize our results in Table.~\ref{aba:tbl2}. The interesting cosmological solutions quoted in this theory are all stemmed from pure geometrical effects.



\section*{Acknowledgments}
The work of MBL is supported by the Portuguese Agency Funda\c{c}\~{a}o para a Ci\^{e}ncia e Tecnologia
through an Investigador FCT Research contract, with reference IF/01442/2013/ CP1196/CT0001. C.-Y.C. and P.C. are supported by Taiwan National Science Council under Project No. NSC 97-2112-M-002-026-MY3 and by Taiwan’s National Center for Theoretical Sciences (NCTS). P.C. is in addition supported by US Department of Energy under Contract No. DE-AC03-76SF00515.


\end{document}